\documentclass[12pt, twoside]{article} 
\usepackage{amssymb}
\usepackage{amsmath} 
\usepackage{amscd}
\usepackage[all]{xy} 
\usepackage{amsfonts} 
\usepackage{amsthm} 
\usepackage{eucal}
\usepackage{latexsym} 
\theoremstyle{definition} 
\newtheorem{defn}{Definition}[section] 
\theoremstyle{plain} 
\newtheorem{thm}{Theorem}[section] 
\newtheorem{lem}{Lemma}[section] 
\newtheorem{cor}{Corollary}[section]
\theoremstyle{remark} 
\newtheorem{exmp}{Example}[section]
\newtheorem{rem}{Remark}[section] 
\newtheorem{obs}{Observation}
\newtheorem{ax}{Axiom}
\newtheorem{as}{Assertion}
\newcommand{\field}[1]{\mathbb{#1}} 
\newcommand{\A}{\field{A}} 
\newcommand{\B}{\field{B}}
\newcommand{\F}{\field{F}} 
\newcommand{\R}{\field{R}} 
\newcommand{\C}{\field{C}}

\newcommand{\N}{\field{N}}
\newcommand{\I}{\field{I}} 
\newcommand{\HH}{\field{H}}
\newcommand{\V}{\field{V}}
\newcommand{\X}{\field{X}}
\newcommand{\bo}[1]{\boldsymbol{#1}} 
\newcommand{\mf}[1]{\mathsf{#1}} 
\newcommand{\mc}[1]{\mathcal{#1}} 
\newcommand{\mb}[1]{\mathbf{#1}}
\newcommand{\mk}[1]{\mathfrak{#1}} 
\begin{document} 
\title{GR-friendly description of quantum systems} 
\author{\emph{International} \emph{Journal} \emph{of} \emph{Theoretical} \emph{Physics}\\ $\bo{47 (2)}$ (2008) 492-510\\ 
DOI: 10.1007/s10773-007-9474-3\\ \\Vladimir Trifonov\\American Mathematical Society}
\date{}\maketitle 
\begin{abstract} 
We present an axiomatic modification of quantum mechanics with a \emph{possible} \emph{worlds} semantics capable of \emph{predicting} essential ``nonquantum'' features of an observable universe model  - the \emph{topology} and \emph{dimensionality} of spacetime, the \emph{existence}, the \emph{signature} and a \emph{specific} \emph{form} of a metric on it, and a set of naturally preferred directions (\emph{vistas}) in spacetime unrelated to its metric properties.   
\end{abstract}
\section{Introduction} \label{Intr}  
The technical purpose of the paper is to provide a formal definition of the notion of an \emph{observer} and related constructs in order to make the description of quantum systems more compatible with the kinematic structure of general relativity (GR). We consider the following kinematic axioms of GR: 
\begin{ax}  Spacetime of the universe is a smooth manifold. \end{ax} 
\begin{ax} The dimensionality of spacetime is four. \end{ax} 
\begin{ax} Spacetime is equipped with a lorentzian metric. \end{ax} 
The fact that these \emph{still} are axioms is somewhat troublesome, at least to the author, who would rather see them as corollaries of a single assertion.
\par We shall consider changes to the kinematic, dynamic and semantic structure of quaternionic quantum mechanics (QQM, \cite{Adl95}) that deal with notions pertaining to the above axioms and the following assertion: 
\begin{as} The logic of the observer is bivalent boolean. \end{as} 
We shall show that (technically accurate versions of) the statements of the axioms are derivable from (a technically accurate version of) this assertion. We start with three simple observations. 
\begin{obs} \label{Kinematics} In QQM the quaternionic hilbert space $\V$ contains a natural principal bundle, $(\mc{V}_{\HH}^{\circ})$, with the following components: \begin{enumerate} 
\item The total space is the set $\mc{V}_{\HH}^{\circ} := \V \setminus \{\bo{0}\}$ of nonzero vectors of $\V$, with the natural manifold structure canonically generated by the linear structure of $\V$.  
\item The base space is a quaternionic projective space $\mc{P}\mc{V}_{\HH}^{\circ}$ whose points are quaternionic rays in $\V$. 
\item The standard fiber is the set $\mc{H}^{\circ} = \HH \setminus \{\bo{0}\}$ of nonzero quaternions which is a four dimensional manifold and a lie group $\mc{H}^{\circ} \cong SU(2)\times \R^+$. 
\item The structure group is also $\mc{H}^{\circ}$. It acts on $\mc{V}_{\HH}^{\circ}$ from the left, and for each $\psi \in \mc{V}_{\HH}^{\circ}$ its orbit (the ray through $\psi$) is a copy of $\mc{H}^{\circ}$ via the fiber diffeomorphism. 
\end{enumerate} We shall refer to this bundle as the  \emph{hyperquantum} \emph{bundle} \emph{over} $\V$. The recent discovery \cite{Tri07} of natural relativistic structure on $\mc{H}^{\circ}$  turns each fiber of $(\mc{V}_{\HH}^{\circ})$ into a lorentzian manifold.  This principal bundle is a  generalization of a principal bundle $(\mc{V}^{\circ}_{\C})$ used in geometric quantum mechanics (GQM), the \emph{quantum} \emph{bundle} \emph{over} a \emph{complex} hilbert space $\V$ (see \cite{Kib79}, \cite{Sch96} and references therein). The hyperquantum bundle forms the kinematic foundation of the modification. \end{obs}
\begin{obs} \label{Dynamics} The total space $\mc{V}_{\HH}^{\circ}$ has the structure of a hyperk\"{a}hler manifold with the riemannian metric and symplectic forms induced by the decomposition of values of the quaternionic hermitian form on $\V$ in the canonical basis $(\bo{1}, \bo{i}, \bo{j}, \bo{k})$ of quaternions. Hence vector fields and flows on $\mc{V}_{\HH}^{\circ}$ are subject to the hyperhamiltonian formalism \cite{GM02}, roughly a superposition of three hamiltonian evolutions on $\mc{V}_{\HH}^{\circ}$, on which the dynamics of the modification is based. \end{obs}
\begin{obs} \label{Semantics}  The observer theory sketched in \cite{Tri95} supplies material for a rigorous definition of an \emph{observer} and  \emph{perceptible} analogues of standard physical constructs such as \emph{time}, \emph{spacetime} and a \emph{dynamical} \emph{system}, and study the dependence of their properties on the type of logic of the observer. The \emph{perceptible} \emph{spacetime} acquires a group structure, and a \emph{perceptible} \emph{dynamical} \emph{system} becomes a set with a left action of a monoid. The main result of \cite{Tri95} asserts that if the logic of an observer is bivalent boolean then the perceptible spacetime is isomorphic to the lie group of nonzero quaternions, $\mc{H}^{\circ}$, and a perceptible dynamical system is a set with a left action of $\mc{H}^{\circ}$. A modest category-theoretic generalization of this schema provides a semantic foundation of the modification. \end{obs}  
In this paper we restate and combine these observations in a coherent way. It should be stressed that the proposed modification is not a complete theory because the treatment of certain important aspects related to quantum measurement and probability is too sketchy. Here our primary objective is \emph{complete} \emph{unambiguity}, i.~e., axiomatic consistency rather than completeness. The paper is organized as follows.  
\par \emph{Section} \ref{TechPr} - Basic constructs and notations. The technical material is given very selectively, the purpose being an introduction of notational conventions rather than education. For example, the reader is assumed to have some familiarity with birkhoff categories and hyperk\"{a}hler geometry. Since this paper utilizes some conventional structures ($\F$-algebras) in an unconventional way, thorough knowledge of \cite{Tri07} is essential for constructive reading. In particular, the notions of a \emph{principal} \emph{inner} \emph{product} on an $\F$-algebra and a \emph{principal} \emph{metric} on a unital algebra, introduced in \cite{Tri07}, are crucial to our treatment. $\N$ denotes natural numbers and zero, $\C$ an $\R$ are the fields of complex and real numbers, respectively, and $\R$ is assumed taken with its standard linear order and euclidean topology. Small Greek indices, $\alpha, \beta, \gamma$ and small Latin indices $p, q$ \emph{always} run $0$ to $3$ and $1$ to $3$, respectively. The set of $[\begin{smallmatrix} n\\m \end{smallmatrix}]$-tensor fields on a smooth  manifold $\mc{M}$ is denoted $\mc{M}[\begin{smallmatrix} n\\m \end{smallmatrix}]$. 
\\ \emph{Section} \ref{HHF} - A slightly nonstandard description of hyperk\"{a}her manifolds and basic notions associated with quaternionic hilbert spaces and quaternionic regular maps.
\\ \emph{Section} \ref{Sem} - The semantics of the modification. We introduce the notions of an \emph{experient} and  a \emph{reality}. The reader is warned that we shall neither discuss the philosophical issues involved, nor use the terminology and notation that usually accompany them (see for example \cite{BFG97}, \cite{Cha96}). Ours is a simpler and more pragmatic task -- to \emph{formally} redefine certain technical constructs of physics in terms of \emph{elementary} \emph{experiences} of an observer, considered as \emph{primitive} \emph{entities}. 
\\ \emph{Section} \ref{FOBS} - We study a particular species of experients called $\F$-\emph{observers} and the associated notions of a \emph{temporal} \emph{reality} and a \emph{phenomenon}. 
\\ \emph{Section} \ref{Obs} - A specialization of some of the above notions, namely an \emph{observer} and a \emph{robust} \emph{reality}. 
\\ \emph{Section} \ref{DS} - We define the notion of a \emph{dynamical} \emph{system}, its \emph{evolution} and its \emph{perceptibles}. The important notion of \emph{propensity} is defined.
\\ \emph{Section} \ref{Cosm} - A special case of a robust reality, a \emph{cosmology}. This Section contains one of the central results of the paper, namely the essential uniqueness of the cosmology of the observer. We compute several  characteristics of the cosmology such as topology and dimensionality of its spacetime, as well as properties of its metric and naturally preferred directions in the spacetime. 
\\ \emph{Section} \ref{PHSYS} - We define \emph{physical} \emph{systems}, \emph{observables} and their \emph{measurements}.  It is shown that standard quantum systems of complex quantum mechanics correspond to a \emph{degenerate}, in a strictly defined sense, kind of physical systems, and the notions of a measurement and propensity seem to reflect certain aspects of quantum measurement and probability, respectively. 
\\ \emph{Section} \ref{SUM} - We conclude the paper with an informal summary of the results. 
\section{Technical preliminaries} \label{TechPr} 
\begin{defn} A \emph{signature}, $\Sigma$, is an ordered pair $(\mf{S}, \mf{s})$, where $\mf{S}$ is a set of \emph{elementary} \emph{symbols} and $\mf{s} : \mf{S} \to \N$ is the \emph{arity} \emph{map}, assigning to each elementary symbol $s \in \mf{S}$ a natural number $\mf{s}(s)$, called the \emph{arity} of $s$. \end{defn} 
\begin{defn} For a category $\mc{E}$ with products and coproducts and a signature $\Sigma = (\mf{S}, \mf{s})$, an endofunctor $\Gamma : \mc{E} \to \mc{E}$  is called a  $\Sigma$-\emph{functor} (\emph{on} $\mc{E}$) if for each $\mc{E}$-object $\mf{A}$,  
\begin{equation} \Gamma(\mf{A}) = \coprod_{s \in \mf{S}} \mf{A}^{\mf{s}(s)} , \end{equation} where $\coprod$ denotes coproduct of $\mc{E}$-objects, and $\mf{A}^{\mf{s}(s)}$ is a product of $\mf{s}(s)$ copies of $\mf{A}$. \end{defn} 
\begin{defn} Given an endofunctor $\Gamma : \mc{E} \to \mc{E}$ on a category $\mc{E}$, an \emph{algebra}, $\A$, \emph{for} $\Gamma$ is an ordered pair $(\mf{A},  \mf{a})$, where $\mf{A}$ is an $\mc{E}$-object, called the \emph{carrier}, and $ \mf{a} : \Gamma(\mf{A}) \to \mf{A}$ is an $\mc{E}$-arrow, called the \emph{structure} \emph{map} of the algebra. Let $\A = (\mf{A}, \mf{a})$ and $\B = (\mf{B}, \mf{b})$ be algebras for $\Gamma$. A $\Gamma$-\emph{morphism}, $\A \to \B$, is a map $f: \mf{A} \to \mf{B}$, such that the following diagram commutes:
\begin{equation} \begin{CD} \Gamma(\mf{A}) @> \Gamma(f) >> \Gamma(\mf{B}) \\ @V  \mf{a} VV @VV \mf{b} V\\ 
\mf{A} @> f >> \mf{B} \end{CD} \quad . \end{equation} \end{defn} 
\begin{exmp} Let $\F$ be a field, and $\mf{S} = \{\bo{0}, +\} \cup \F$ and $\mf{s}(\bo{0}) = 0$, $\mf{s}(+) = 2$, $\mf{s}(r) = 1, \forall r \in \F$. A \emph{vector} \emph{space}, $V = (\mf{V}, \mf{v})$, \emph{over} \emph{a} \emph{field} $\F$ is an algebra for the $\Sigma$-functor $\Gamma : \bo{Set} \to \bo{Set}$ on the category of sets for the signature $\Sigma = (\mf{S}, \mf{s})$.  \end{exmp} 
\begin{rem} Dually, \emph{coalgebras} \emph{for} \emph{an} \emph{endofunctor} $\Gamma : \mc{E} \to \mc{E}$ are defined by reversal of structure maps. Algebras and coalgebras for an endofunctor $\Gamma$ and their $\Gamma$-morphisms form categories denoted $\mc{E}^{\Gamma}$ and  $\mc{E}_{\Gamma}$, respectively, (see, e.~g., \cite{Hug01}).  \end{rem} 
\begin{exmp} A monoid is an example of an algebra, $M = (\mf{M}, \mf{m})$, for a $\Sigma$-functor on $\bo{Set}$ with $\mf{S} = \{\imath, \ast\}, \mf{s}(\imath) = 0, \mf{s}(\ast) = 2$. As with every $\Sigma$-functor, the structure map $\mf{m}$ can be split into the constituents, giving the more familiar notation $(\mf{M}, \imath, \ast)$, where $\imath$ is understood as a preferred element, the identity of $M$, and $\ast$ is a binary operation on $\mf{M}$.  \end{exmp} 
\begin{defn} An endofunctor $\Gamma$ on $\mc{E}$ is called a \emph{monad} if there exist two natural transformations, $\flat : id(\mc{E}) \to \Gamma$ and $\natural : \Gamma^2 \to \Gamma$ such that the following diagrams commute: \begin{equation}
\xymatrix{\Gamma^3 \ar[d]_{\natural\Gamma} \ar[r]^{\Gamma\natural} & \Gamma^2 \ar [d]^\natural \\ 
\Gamma^2 \ar[r]^{\natural \Gamma} & \Gamma}  \quad \xymatrix{\Gamma \ar @{=}[dr] \ar [r]^{\Gamma\flat} & \Gamma^2 \ar [d]|\natural & \ar [l]_{\flat\Gamma} \Gamma \ar @{=}[dl]\\& \Gamma &} , \end{equation} where $\Gamma^n$ denotes $n$ iterations of the functor. \end{defn} 
\begin{defn} The \emph{category} \emph{of} \emph{algebras}, $\bar{\mc{E}}^{\Gamma}$, \emph{for} \emph{the} \emph{monad} $\Gamma : \mc{E} \to \mc{E}$ is a subcategory of $\mc{E}^{\Gamma}$ such that the following diagrams commute for each object $\A =(\mf{A},  \mf{a})$ of $\bar{\mc{E}}^{\Gamma}$: \begin{equation}
\xymatrix{\Gamma^2(\A) \ar[d]_{\natural(\A)} \ar[r]^ {\Gamma(\mf{a})} & \Gamma(\A) \ar [d]^ {\mf{a}} \\ 
\Gamma(\A) \ar[r]^ {\mf{a}} & \A}  \quad \xymatrix{\A \ar [dr]_ {id(\A)} \ar [r]^ {\flat(\A)} & \Gamma(\A) \ar [d]^ {\mf{a}} \\& \A} \end{equation} \end{defn} 
\begin{exmp} \label{PU} For each monoid $M = (\mf{M}, \mf{m})$, an endofunctor $\Gamma$ on $\bo{Set}$ sending a set $\mf{X}$ to the set $\mf{M} \times \mf{X}$ is a monad. An object, $\X = (\mf{X}, \mf{x})$, in the category of algebras, $M\bo{Set}$, for this monad is a set, $\mf{X}$, with a left action, $\mf{x}$, of the monoid $M$, also referred to as an $M$-\emph{set}. For each $a \in \mf{M}$, we can define a map $\mf{x}_a : \mf{X} \to \mf{X}$ by $\mf{x}_a(x) := \mf{x}(a, x), \forall x \in \mf{X}$. The arrows $(\mf{X}, \mf{x}) \to (\mf{X}^\prime, \mf{x}^\prime)$ are \emph{equivariant} (i.~e., preserving the action) functions $f : \mf{X} \to \mf{X}^\prime$, making the following diagram commute: 
\begin{equation} \begin{CD} \mf{X} @> f >> \mf{X}^\prime \\ @V \mf{x}_a VV @VV \mf{x}^\prime_a V\\ 
\mf{X} @> f >> \mf{X}^\prime \end{CD} \quad . \end{equation} The category of $M$-sets is of utmost importance in our exposition. 
 \end{exmp} 
\begin{exmp} An $\F$-\emph{algebra} (a linear algebra over the field $\F$) is an example of an algebra for a $\Sigma$-functor on $\bo{Set}$ with $\mf{S} = \{\bo{0}, +, \cdot\} \cup \F$ and $\mf{s}(\bo{0}) = 0, \mf{s}(+) = 2, \mf{s}(\cdot) = 2, \mf{s}(r) = 1, \forall r \in \F$. For an unconventional description of $\F$-algebras, as vector spaces over $\F$ with $[\begin{smallmatrix} 1\\2 \end{smallmatrix}]$-tensors on them (\emph{structure} \emph{tensors}), crucial to our technical setup, see \cite{Tri07}.  \end{exmp} 
\begin{defn} A complete regularly co-well-powered category with regular epi-mono factorizations is called \emph{birkhoff} if it has enough regular projectives. \end{defn} 
\begin{rem} For a detailed view of birkhoff categories see \cite{Hug01}. \end{rem}
\begin{defn} A full subcategory of a Birkhoff category is called a \emph{birkhoff} \emph{variety} if it is closed under products, subobjects and quotients. \end{defn} 
\begin{defn} An endofunctor $\Gamma : \mc{E} \to \mc{E}$ is called a \emph{varietor} if it preserves regular epis, and the forgetful functor $U : \mc{E}^\Gamma \to \mc{E}$ has a left adjoint. \end{defn}
A real vector space $V$ induces a natural manifold structure on its carrier. This manifold which we denote $\mc{V}$, is  referred to as the \emph{linear} \emph{manifold} \emph{canonically} \emph{generated} \emph{by} $V$.  Since $V$ and $\mc{V}$ have the same carrier, there is a bijection $\mc{J}_V : \mc{V} \to V$. We use the normal ($a, u, ...$) and bold ($\bo{a}, \bo{u}...$) fonts,  to denote the elements of $\mc{V}$ and $V$, respectively, e.~g., $\mc{J}_V(a) = \bo{a}$. The tangent space $T_a\mc{V}$ is identified with $V$ at each point $a \in \mc{V}$ via an isomorphism $\mc{J}^*_a : T_a\mc{V} \to V$ sending a tangent vector to the curve $\mu : \R \to \mc{V}, \mu(t) =  a + tu$, at the point $\mu(0) = a \in \mc{V}$, to the vector $\bo{u} \in V$, with the ``total'' map $\mc{J}_V^* : T\mc{V} \to V$. The set of nonzero vectors of $V$ constitutes a submanifold of $\mc{V}$, referred to as the \emph{punctured} \emph{manifold} (\emph{of} $V$), denoted $\mc{V}^{\circ}$. 
\begin{defn} For a real vector space $V$ and a linear map $\bo{F} : V \to V$, a vector field $\bo{f} : \mc{V} \to T\mc{V}$ on $\mc{V}$, such that the following diagram commutes \begin{equation} \label{VM} \begin{CD} \mc{V} @> \bo{f} >> T\mc{V} \\ @V {\mc{J}_V} VV @VV {\mc{J}_V^*} V\\ 
V @> \bo{F} >> V \end{CD} \quad , \end{equation} is called the \emph{vector} \emph{field} \emph{canonically} \emph{generated} by $\bo{F}$. \end{defn} 
\begin{rem} In particular, for a unital algebra $\A$, the linear manifold canonically generated by its underlying vector space $A$ is denoted by $\mc{A}$ and the punctured manifold by $\mc{A}^{\circ}$. \end{rem}
\begin{defn} For real vector spaces $U$, $V$, and a linear map $\bo{F} : U \to V$, the map $F : \mc{U} \to \mc{V}$, such that the following diagram commutes:
\begin{equation} \label{UV} \begin{CD} \mc{U} @> F >> \mc{V} \\ @V {\mc{J}_U} VV @VV {\mc{J}_V} V \\ 
U @> \bo{F} >> V \end{CD} \quad , \end{equation} is called the \emph{linear} \emph{induction} of $\bo{F}$. \end{defn} 
\section{Quaternionic maps and hyperhamiltonian vector fields} \label{HHF} 
\begin{defn} For the quaternion algebra $\HH = (H, \bo{\mf{H}})$ a basis on the vector space $H$ is called \emph{canonical} if the components of the structure tensor $\bo{\mf{H}}$ are given by the entries of the following matrices
\begin{multline} \mf{H}^0_{\alpha \beta} = 
\begin{pmatrix} 1&0&0&0\\0&-1&0&0\\0&0&-1&0\\0&0&0&-1 \end{pmatrix},\ 
\mf{H}^1_{\alpha \beta} = \begin{pmatrix} 0&1&0&0\\1&0&0&0\\0&0&0&1\\0&0&-
1&0 \end{pmatrix}, \\ \mf{H}^2_{\alpha \beta} = \begin{pmatrix} 
0&0&1&0\\0&0&0&-1\\1&0&0&0\\0&1&0&0 \end{pmatrix},\ \mf{H}^3_{\alpha 
\beta} = \begin{pmatrix} 0&0&0&1\\0&0&1&0\\0&-1&0&0\\1&0&0&0 
\end{pmatrix}. \end{multline} \end{defn}
\begin{rem} The set $\bo{\{i\}}$ of canonical bases $(\bo{i}_{\beta})$ is parametrized by the elements of $SO(3)$. We shall refer to a bijection $\Xi : SO(3) \to \bo{\{i\}}$ as an \emph{array}. \end{rem}
\begin{rem} For a set $\mf{X}$, the values of a map $f : \mf{X} \to \HH$ can be decomposed in a canonical basis $(\bo{i}_{\beta})$, producing an ordered quadruple $(f_{\beta})$ of real valued maps on $\mf{X}$ which we call the \emph{constituents} of $f$ in the basis $(\bo{i}_{\beta})$. We shall be careful to distinguish these from components of tensorial objects. \end{rem}
Let $\mc{M} = (M, \bo{g}, (\tilde{\bo{\omega}}^p))$ be a hyperk\"{a}hler manifold. Then, for an array $\Xi : SO(3) \to \bo{\{i\}}$, we can consider the ordered quadruple $(\bo{g}, \tilde{\bo{\omega}}^1, \tilde{\bo{\omega}}^2, \tilde{\bo{\omega}}^3))$ as  constituents $(\tilde{\bo{\omega}}^{\beta})$, where $\tilde{\bo{\omega}}^0 := \bo{g}$, of a quaternion valued map $\tilde{\bo{\omega}} : T\mc{M} \times T\mc{M} \to \HH$ in the standard basis $\Xi(\bo{1})$, assigning to each ordered pair $(\bo{u}, \bo{v})$ of tangent vectors at each point $\phi \in \mc{M}$ a quaternion 
$\tilde{\bo{\omega}}^{\beta}(\bo{u}, \bo{v})\bo{i}_{\beta}$, which allows us to define the constituents $\tilde{\bo{\omega}}^p$ in every canonical basis $(\bo{i}_{\beta})$. Similarly, we can assign the ordered quadruple $(\mc{I}_{\beta})$, where $\mc{I}_0$ is the identity map on $T\mc{M}$, and $\mc{I}_p$ is the complex structure corresponding to $\tilde{\bo{\omega}}^p$, to each canonical basis $(\bo{i}_{\beta})$, as constituents of the hypercomplex structure $\bo{\mc{I}}$. 
\begin{defn} For an array $\Xi$ and a hyperk\"{a}hler manifold $\mc{M}$, the ordered pair $\bo{\Omega} : (\tilde{\bo{\omega}}, \bo{\mc{I}})$ is called a \emph{hyperk\"{a}hler} \emph{structure} on $\mc{M}$ \emph{generated} \emph{by} $\Xi$. For a canonical basis $(\bo{i}_{\beta})$ the maps $\tilde{\bo{\omega}}^p$ and $\mc{I}_p$ are called the \emph{symplectic} and \emph{complex} \emph{constituents} of $\bo{\Omega}$, respectively, \emph{in} the basis $(\bo{i}_{\beta})$. \end{defn}   
\begin{defn} A left module $\mf{V}$ over $\HH$ is called a \emph{(left)} \emph{quaternionic} \emph{vector} \emph{space}. \end{defn} 
\begin{rem} A \emph{right} \emph{quaternionic} \emph{vector} \emph{space} is defined similarly, as well as the right versions of constructs based on it. \end{rem}
\begin{defn} A map $\widehat{\bo{F}} : \mf{V} \to \mf{V}$ is called a \emph{(left)} \emph{quaternion} \emph{linear} \emph{operator}, if 
\begin{displaymath} \widehat{\bo{F}}(\bo{a}\bo{\phi}) = \bo{a}\widehat{\bo{F}}(\bo{\phi}), \quad \forall \bo{\phi} \in \mf{V}, \quad \forall \bo{a} \in \HH . \end{displaymath} \end{defn} 
\begin{defn} A \emph{(left)} \emph{quaternionic}  \emph{hilbert} \emph{space}, $\V$, is a quaternionic vector space, $\mf{V}$, together with a map  $\langle\cdot\mid\cdot\rangle : \mf{V} \times \mf{V} \to \HH$, called a \emph{quaternionic} \emph{hermitian} \emph{inner} \emph{product}, such that  
\begin{multline*} \quad \langle\bo{\phi} \mid \bo{\psi} + \bo{\xi}\rangle = \langle\bo{\phi} \mid \bo{\psi}\rangle + \langle\bo{\phi} \mid \bo{\xi}\rangle, \\ \langle\bo{\phi} \mid \bo{\psi}\rangle = \overline{\langle\bo{\psi} \mid \bo{\phi}\rangle}, \quad \langle\bo{a}\bo{\phi} \mid \bo{\psi}\rangle = \bo{a}\langle\bo{\phi} \mid \bo{\psi}\rangle , \\ 
\parallel\bo{\phi}\parallel^2 := \langle\bo{\phi} \mid \bo{\phi}\rangle \in \R,  \parallel\bo{\phi}\parallel^2 > 0, \quad \forall \bo{\phi} \neq \bo{0}, \\ \forall \bo{\phi}, \bo{\psi}, \bo{\xi} \in \mf{V} , \forall \bo{a} \in \HH , \end{multline*} 
and the diagonal $\parallel\cdot\parallel$ induces a topology on $\mf{V}$, relative to which $\mf{V}$ is separable and complete. \end{defn} 
\begin{defn} \label{ADJ} For a quaternionic hilbert space $\V$ and a quaternion linear operator $\widehat{\bo{F}}$ on $\V$, its \emph{quaternionic} \emph{adjoint} (\emph{with} \emph{respect} \emph{to}  $\langle\cdot\mid\cdot\rangle$) is a quaternion linear operator $\widehat{\bo{F}}^{\dagger}$ on $\mf{V}$, such that 
\begin{displaymath} \langle\bo{\phi} \mid \widehat{\bo{F}}(\bo{\psi})\rangle = \langle\widehat{\bo{F}}^{\dagger}(\bo{\phi}) \mid \bo{\psi}\rangle , \quad \forall \bo{\phi}, \bo{\psi} \in \V . \end{displaymath} \end{defn} 
\begin{defn} We refer to a quaternion linear operator $\widehat{\bo{F}}$ as \emph{quaternionic} \emph{(anti)hermitian} if it coincides with (the negative of) its quaternionic adjoint.  \end{defn} 
\begin{rem} Given a basis, $(\bo{e}_j)$, on an $n$ dimensional quaternionic hilbert space $\V$, and a canonical basis, $(\bo{i}_{\beta})$, on $H$,  $\V$ induces a real $4n$ dimensional vector space, $V$, and the latter canonically generates a real linear $4n$ dimensional manifold $\mc{V}_{\HH}$ and the punctured manifold $\mc{V}_{\HH}^{\circ}$. \end{rem}
\begin{defn} For an operator $\bo{\widehat{F}} : \V \to \V$ on a quaternionic hilbert space $\V$ its \emph{expectation} \emph{operator} is a map $\widehat{F} : \V \to \HH$ assigning to each $\bo{\phi} \in \V$ a quaternion $\langle\bo{\phi} \mid \bo{\widehat{F}}(\bo{\phi}) \rangle$. \end{defn} 
\begin{lem} For each quaternionic antihermitian operator, $\bo{\widehat{F}}$, its expectation operator $\widehat{F}$ has the following property:
\begin{equation} \widehat{F}(\bo{\phi}) = \mk{Im}(\widehat{F}(\bo{\phi})), \quad \forall \bo{\phi} \in \V . \end{equation} \end{lem} 
\begin{proof} \begin{equation} \langle\bo{\phi} \mid \bo{\widehat{F}}(\bo{\phi}) \rangle = \langle -\bo{\widehat{F}}(\bo{\phi}) \mid \bo{\phi}\rangle = -\langle \bo{\widehat{F}}(\bo{\phi}) \mid \bo{\phi}\rangle  = -\overline{\langle \bo{\phi} \mid \bo{\widehat{F}}(\bo{\phi})\rangle} . \end{equation} 
Thus, for each $\bo{\phi} \in \V$ the quaternion $\langle\bo{\phi} \mid \bo{\widehat{F}}(\bo{\phi}) \rangle$ coincide with the negative of its adjoint, which means that its real part is zero. \end{proof} 
\begin{defn} For a quaternionic antihermitian operator $\bo{\widehat{F}} :\V \to \V$ the linear induction $F : \mc{V}_{\HH} \to \mc{H}$ of its expectation operator $\widehat{F}$ is called the \emph{expectation} of $\bo{\widehat{F}}$. \end{defn}
\begin{defn} For a quaternion linear operator $\widehat{\bo{F}} : \V \to \V$ a vector field $\bo{f} : \mc{V}_{\HH} \to T\mc{V}_{\HH}$ canonically generated by $-\widehat{\bo{F}}$ is called the \emph{hyperfield} of $\widehat{\bo{F}}$. \end{defn}
\begin{rem} The set of the integral curves of the hyperfield of $\widehat{\bo{F}}$ can be formally represented by the following differential equation
\begin{equation} \label{SCHR} \dot{\psi} = -\widehat{\bo{F}}(\psi) . \end{equation} 
Hyperfields are a generalization of \emph{schr\"{o}dinger} \emph{vector} \emph{fields} of geometric quantum mechanics \cite{Sch96}. Indeed, due to the simple relationship between eigenvalues of hermitian and antihermitian operators in the complex case we can use either kind to represent observables. If we take antihermitian operators, the Schr\"{o}dinger equation has the form \eqref{SCHR}, with $\widehat{\bo{F}} = \widehat{\bo{H}}$, which defines the set of integral curves of the schr\"{o}dinger vector field of the hamiltonian operator $\widehat{\bo{H}}$. \end{rem} 
Given an array $\Xi$, the quaternionic hermitian product $\langle\cdot \mid \cdot \rangle$ induces a map $T\mc{V}_{\HH} \times T\mc{V}_{\HH} \to \HH$ whose  constituents in a canonical basis $(\bo{i}_{\beta})$ can be identified with the symplectic constituents in $(\bo{i}_{\beta})$ of a hyperk\"{a}hler structure $\bo{\Omega}$ on $\mc{V}_{\HH}$ generated by $\Xi$. Similarly, the maps $(\bo{\imath}^{\prime}_{\beta})$ defined by 
\begin{displaymath} \bo{\imath}^{\prime}_{\beta}(\bo{\phi}) := \bo{i}_{\beta}\bo{\phi}, \quad \forall \bo{\phi} \in \V \end{displaymath} 
produce the complex constituents of $\bo{\Omega}$.  
\begin{exmp} The quaternion algebra $\HH$ is a quaternionic vector space. Together with a natural quaternionic hermitian inner product defined by 
\begin{equation} \langle\bo{a} \mid \bo{b} \rangle := \bo{a}\overline{\bo{b}}, \quad \forall \bo{a}, \bo{b} \in \HH , \end{equation} it is a quaternionic hilbert space. Therefore $\mc{H}$ and $\mc{H}^{\circ}$ possess natural hyperk\"{a}hler structures. \end{exmp} 
\begin{defn} Let $\mc{M}$ and $\mc{N}$ be hyperk\"{a}hler manifolds with hyperk\"{a}hler structures $\bo{\Omega}^{\mc{M}}$ and $\bo{\Omega}^{\mc{N}}$, respectively, generated by an array $\Xi$.  A smooth map $f : \mc{M} \to \mc{N}$ is called \emph{quaternionic} (\emph{regular}) if there exists an $SO(3)$ matrix $\bo{\mf{B}}$ such that 
\begin{equation} \label{QR} \sum_{p, q=1}^3\mf{B}_{pq}\mc{I}^{\mc{N}}_q \circ df \circ \mc{I}^{\mc{M}}_p = df , \end{equation} where $\mc{I}^{\mc{N}}_p$ and $\mc{I}^{\mc{N}}_p$ are the complex constituents, in a canonical basis $(\bo{i}_{\beta}) = \Xi(\bo{\mf{B}})$, of the hyperk\"{a}hler structures of $\mc{M}$ and $\mc{N}$, respectively, and $df$ is the differential of $f$. \end{defn}
\begin{rem} Quaternionic maps generalize holomorphic functions of complex analysis (see e.~g., \cite{CL00}). \end{rem} 
\begin{defn} For a quaternionic map $f : \mc{M} \to \mc{H}$, its  \emph{hyperhamiltonian} \emph{vector} \emph{field} is a vector field $\bo{f}$ on $\mc{M}$, such that 
\begin{equation} \label{HHE1} (df_0)(\bo{u}) =  \bo{g}(\bo{f}, \bo{u}) , \quad \forall \bo{u} \in \mc{M}[\begin{smallmatrix} 1\\0 \end{smallmatrix}] , \end{equation} 
where $\bo{g}$ is the riemannian metric on $\mc{M}$, and $df_0$ is the differential of $f_0$. The maps $f$ and $f_0$ are referred to as the \emph{generating} \emph{map} and the \emph{main} \emph{generator} of $\bo{f}$, respectively. \end{defn} 
\begin{rem} Since $f_0$ is invariant under a canonical basis change, so is the definition of a hyperhamiltonian vector field. \end{rem}
\begin{thm} For an array $\Xi$ and a quaternionic hilbert space $\V$ let $\bo{f}$ be the hyperhamiltonian vector field of a quaternionic map $f : \mc{V}_{\HH} \to \mc{H}$, such that $\mk{Im}(f)$ is the expectation of a quaternion antihermitian operator $\bo{\widehat{F}}$ on $\V$. Then $\bo{f}$ is the hyperfield of $\bo{\widehat{F}}$, and for each canonical basis $(\bo{i}_{\beta})$ there exists an ordered triple $(\bo{f}_p)$ of vector fields on $\mc{V}_{\HH}$ such that 
\begin{equation} \label{GMT} \bo{f} = \bo{f}_1 + \bo{f}_2 + \bo{f}_3 , \quad (df_p)(\bo{u}) = \tilde{\bo{\omega}}^p(\bo{f}_p, \bo{u}), \quad \forall \bo{u} \in \mc{V}_{\HH}[\begin{smallmatrix} 1\\0 \end{smallmatrix}] ,  \end{equation} (no summation on $p$) where $df_p$ is the differential of $f_p$, and $\tilde{\bo{\omega}}^p$ is a symplectic constituent of the hyperk\"{a}hler structure on $\mc{V}_{\HH}$ generated by $\Xi$, in the basis $(\bo{i}_{\beta})$. \end{thm} 
\begin{proof} The hyperfield of $\bo{\widehat{F}}$ is clearly a hyperhamiltonian vector field whose generating map is $f$ \cite{Sch96}. It was shown in \cite{GM02} and \cite{MT03} that given a quaternionic map $f$, the conditions \eqref{GMT} are satisfied for a unique vector field $\bo{f}$, and hence the hyperfield of $\bo{\widehat{F}}$ coincides with $\bo{f}$. \end{proof} 
\section{Semantics} \label{Sem} 
The observer theory outlined in \cite{Tri95} describes an observer as an \emph{experient}, as opposed to an \emph{occupant}, of the environment. This can be presented within a framework similar to what is known as the \emph{initial} \emph{algebra} and \emph{final} \emph{coalgebra} approach  to syntax and semantics of formal languages (\cite{Acz97}, \cite{Tur96}). The underlying idea is that perception and comprehension are somehow dual to each other, and each \emph{experience} can be considered as both a \emph{percept} and a \emph{concept}. This idea has its origin in logic and theoretical computer science where similar dualities are considered (a symbol and its meaning, syntax of a formal language and its semantics). We start with a category $\mc{E}$ representing the \emph{totality} \emph{of} \emph{experiences}, and an endofunctor $\Gamma$ on $\mc{E}$ interpreted as the \emph{language} \emph{of} \emph{thought} of the experient (see \cite{Fod75} for a philosophical discussion). The categories of algebras, $\mc{E}^{\Gamma}$, and coalgebras, $\mc{E}_{\Gamma}$, for this endofunctor represent \emph{perception} and \emph{comprehension} of the experient, respectively. If the perception, $\mc{E}^{\Gamma}$, has an initial object, $\I$, the latter serves as the \emph{syntax} of the experient's language of thought. Then an algebra, $\A$, for $\Gamma$ is considered a \emph{paradigm} or a \emph{model} for the language of thought, and the unique arrow $\I \to \A$ as a \emph{meaning} \emph{function} or an \emph{interpretation}. 
\begin{defn} \label{EXP} An \emph{experient}, $\bo{\mc{E}}$, is an ordered pair $(\mc{E}, \Gamma)$, where $\mc{E}$ is a birkhoff category, called the \emph{metauniverse}, and $\Gamma : \mc{E} \to \mc{E}$ is a varietor called the \emph{construer}. The objects and arrows of $\mc{E}$ are called \emph{metaphenomena} and \emph{metalinks}, respectively. $\mc{E}$-elements of metaphenomena are called  \emph{reflexors}. The category of algebras $\mc{E}^{\Gamma}$ is called the \emph{perception} \emph{category}. Objects and arrows of $\mc{E}^{\Gamma}$ are referred to as \emph{paradigms} and \emph{shifts}, respectively. An experient is \emph{coherent} if the metauniverse is a topos. An experient is called \emph{boolean} if the metauniverse is a boolean topos. \end{defn} 
\begin{rem} It follows from the Definition \ref{EXP} that $\mc{E}^{\Gamma}$ is also a birkhoff category (see \cite{Hug01}, p. 125).  \end{rem} 
\begin{rem} Intuitively, metaphenomena are \emph{complex} \emph{percepts} composed out of elementary ones (reflexors). For a coherent experient $\bo{\mc{E}} = (\mc{E}, \Gamma)$, we are to think of the internal logic of $\mc{E}$ as the \emph{metalogic} of the experient. In particular, the metalogic of a boolean experient is boolean.  \end{rem} 
\begin{defn} For a paradigm $\A$ of an experient $\bo{\mc{E}}$, and a natural number $n$, a \emph{reality} \emph{of} \emph{rank} $n$ is an ordered pair $\mc{R} = (\A, \bo{R})$, where $\bo{R}$ is a subobject of $\A^n$. We refer to $\A$ as the \emph{underlying} \emph{paradigm}, and $\bo{R}$ is called the \emph{ontology} of $\mc{R}$. \end{defn}
\begin{rem} Intuitively, a reality of rank $n$ is an $n$-ary relation on a collection of reflexors. From this point on we are interested  exclusively in realities of rank $2$, henceforth referred to simply as \emph{realities}.  \end{rem} 
\begin{defn} An \emph{existence} \emph{mode}, $\bo{\mc{E}}(\mk{V})$, of an experient $\bo{\mc{E}} = (\mc{E}, \Gamma)$ is an ordered pair $(\bo{\mc{E}}, \mk{V})$, where $\mk{V}$ is a birkhoff variety of $\mc{E}^{\Gamma}$. A paradigm $\A$ such that it is (not) a $\mk{V}$-object is called  \emph{(non)existent} \emph{with} \emph{respect} \emph{to} $\bo{\mc{E}}(\mk{V})$. A shift $\A \to \B$ such that it is (not) a $\mk{V}$-arrow is called  \emph{(im)possible} \emph{with} \emph{respect} \emph{to} $\bo{\mc{E}}(\mk{V})$. \end{defn} 
\begin{rem} It should be noted that we thus consider every paradigm $\A$ of an experient also a paradigm of the experient in a certain existence mode, although $\A$ may be nonexistent with respect to the latter. Given an existence mode, $\bo{\mc{E}}(\mk{V})$, of an experient $\bo{\mc{E}}$, we say that the experient \emph{is} \emph{in} \emph{the} \emph{mode} $\mk{V}$, or that the experient \emph{is} the $\mk{V}$-\emph{observer}. \end{rem} 
\section{$\F$-observers} \label{FOBS} Different existence modes possess different amounts of structure to allow for a definition of notions that we normally associate with observers. We shall focus on a class of boolean experients whose metauniverse is $\bo{Set}$, and the construer is a $\Sigma$-functor for the following signature: \begin{multline} \label{FUNCTOR} \Sigma =(\mf{S}, \mf{s}); \quad \mf{S} = \{+, \cdot, \bo{0}\} \cup \F, \quad \mf{s}(+) = 2, \\ 
\mf{s}(\cdot) = 2, \quad \mf{s}(\bo{0}) = 0, \quad \mf{s}(r) = 1, \quad \forall r \in \F , \end{multline} where $\F$ is a field. They are introduced implicitly in \cite{Tri95}, where a particular existence mode, an $\mf{Alg}\{\F\}$-observer, where $\mf{Alg}\{\F\}$ is the category of $\F$-algebras, is studied. $\mf{Alg}\{\F\}$-observers are sufficiently fine structured to define the fundamental notion of a \emph{temporal} \emph{reality}, and at the same time they are relatively simple: their metaphenomena and reflexors are sets and set elements, respectively. For the rest of the paper we shall deal exclusively with $\mf{Alg}\{\F\}$-observers, to whom we henceforth refer simply as $\F$-\emph{observers}.  
\begin{defn} For an existent paradigm $\A$ of an $\F$-observer the underlying vector space $A$ is called the \emph{sensory} \emph{domain} of $\A$, with the principal inner products of $\A$ referred to as \emph{sensory} \emph{forms} of the paradigm. A basis on $A$ is called a \emph{sensory} \emph{basis} of the paradigm. The dual vector space $A^*$ is called the \emph{ether} \emph{domain} of $\A$, with the elements called \emph{ether} \emph{forms}. The \emph{motor} \emph{domain} of $\A$ is the multiplicative subgroupoid, $M_{\A} = (\mf{M}_{\A}, \imath, \ast)$, of $\A$, generated by the set of nonzero reflexors of the paradigm $\A$. We refer to elements of $M_{\A}$ as \emph{effectors}. An existent finite dimensional paradigm, $\A$, is called \emph{rational} if its motor domain is a monoid; otherwise $\A$ is called \emph{irrational}. A paradigm is called \emph{transient} if it is neither rational nor irrational. The set $\mc{A}^{\bullet}$ of invertible reflexors of a rational paradigm $\A$ is called the \emph{perception} \emph{domain} of $\A$. \end{defn} 
\begin{rem} For a rational paradigm $\A$ of an $\F$-observer, the perception domain $\mc{A}^{\bullet}$ is a group, and the ontology, $\bo{R}$, of a reality $\mc{R} = (\A, \bo{R})$ is a binary relation on the carrier of $\A$. \end{rem} 
\begin{defn} A reality of an $\F$-observer is called \emph{immanent} if its underlying paradigm is existent and its ontology is a partial order. Otherwise the reality is called \emph{transcendent}. \end{defn}
\begin{defn} For an $\F$-observer, an immanent reality $(\F, \preceq)$ is called a \emph{temporal} \emph{template} if $\F$ is an ordered field with a complete linear order $\preceq$. \end{defn} 
\begin{rem} Not every existence mode has a temporal template. For instance, complex numbers, $\C$, do not admit linear orders (\cite{Kur73}, p. 304). Therefore the $\C$-observer has no temporal templates.  \end{rem} 
\begin{defn} For a rational paradigm $\A$ of an $\F$-observer the $\A$-\emph{universe} is the category of $M_{\A}$-sets, $M_{\A}\bo{Set}$ (see Example \ref{PU}), with objects and arrows called $\A$-\emph{phenomena} and $\A$-\emph{links}, respectively. For an $\A$-phenomenon $\X = (\mf{X}, \mf{x})$, the metaphenomenon $\mf{X}$ is called the \emph{propensity} \emph{realm}, with elements called \emph{propensity} \emph{modes} of $\X$. Each metalink $\mf{X} \to A$, where $A$ is the carrier of $\A$, is referred to as an \emph{attribute} of $\X$. \end{defn} 
\begin{rem} We refer to the internal logic of the topos $M_{\A}\bo{Set}$ as the \emph{operational} \emph{logic} of the $\F$-observer \emph{with} \emph{respect} \emph{to} $\A$, or as the \emph{logic} of the $\A$-universe.  \end{rem} 
\begin{defn} For each $\A$-phenomenon $\X = (\mf{X}, \mf{x})$, the $\mc{A}^{\bullet}$-set $(\mf{X}, \bar{\mf{x}})$, where $\bar{\mf{x}} : \mc{A}^{\bullet} \times \mf{X} \to \mf{X}$ is the restriction of $\mf{x}$ to $\mc{A}^{\bullet}$, is called the \emph{perceptible} \emph{part} of $\X$. For each $\phi \in \mf{X}$ the orbits, $W_{\phi}$ and $\overline{W}_{\phi}$, of $\phi$, with respect to the actions of $M_\A$ and $\mc{A}^{\bullet}$, respectively, are called an \emph{existence} \emph{mode} and a \emph{presence} \emph{mode} of $\X$, respectively. \end{defn} 
\begin{rem} In other words, $W_{\phi}$ and $\overline{W}_{\phi}$ are the sets 
\begin{multline} W_{\phi} = \{\psi \in \mf{X} \quad : \quad \psi = \mf{x}(\xi, \phi), \quad \forall \xi \in M_\A \} , \\ 
\overline{W}_{\phi} = \{\psi \in \mf{X} \quad : \quad \psi = \bar{\mf{x}}(\xi, \phi), \quad \forall \xi \in \mc{A}^{\bullet} \} . \end{multline} Intuitively, each presence mode of  an $\A$-phenomenon is the perceptible part of one of its existence modes. \end{rem} 
\begin{defn} An $\A$-phenomenon $\X$, together with a map $\sigma : \overline{W} \to \mc{A}^{\bullet}$ for each presence mode $\overline{W}$, is called a \emph{stable} \emph{phenomenon} if each $\sigma$ is a bijection. We refer to $\sigma$ as a \emph{proper} \emph{view} of $\overline{W}$. For a stable phenomenon, the propensity modes $\phi \in (W_{\phi} \setminus) \overline{W}_{\phi}$ are called \emph{(im)perceptible}. \end{defn} 
\begin{defn} For an $\F$-observer, a rational paradigm $\A$ is \emph{consistent} if the $\A$-universe is a boolean topos; a consistent paradigm of maximal dimensionality is called a \emph{home} paradigm of the $\F$-observer. \end{defn} 
\begin{rem} Intuitively, the consistency condition requires the logic of the $\A$-universe to match the logic of the metauniverse (the metalogic of the $\F$-observer). $\F$-observers without home paradigms, referred to as \emph{Wanderers}, may or may not be of interest, but our main concern will be precisely with home paradigms, and more specifically with home paradigms of the $\R$-observer and their $\A$-universes, due to the following result.  \end{rem}
\begin{thm}[Trifonov, 1995] Every home paradigm of the $\R$-observer is isomorphic to the quaternion algebra $\HH$ with a family of minkowski sensory forms. \end{thm} 
\begin{defn} For an $\F$-observer, a reality $\mc{R} = (\A, \bo{R})$ is called \emph{stable} if it is immanent and $\A$ is a rational paradigm; otherwise the reality is called \emph{unstable} (or \emph{virtual}). \end{defn} 
\begin{rem} Given a stable realty $\mc{R} = (\A, \bo{R})$ we refer to the logic of the $\A$-universe also as the \emph{logic} \emph{of} $\mc{R}$. \end{rem}
\begin{defn} For an $\F$-observer with a temporal template $(\F, \preceq)$, a reality $\mc{R} = (\A, \bo{R})$, together with an $\F$-valued map $\mf{T} : \A \to \F$, is called a \emph{temporal} \emph{reality} if it is stable and  
\begin{equation} a \bo{R} b \iff (\mf{T}(a) \preceq \mf{T}(b) \wedge a \neq b) \vee (a = b), \forall a, b \in \A . \end{equation} Otherwise the reality is called \emph{atemporal}.  The ontology of a temporal reality is called its \emph{temporal} \emph{order}, and the map $\mf{T}$ is referred to as the \emph{global} \emph{time} of the reality. The \emph{perceptible} \emph{time}  of $\mc{R}$ is the restriction, $\mc{T} : \mc{A}^{\bullet} \to \F$, of $\mf{T}$ to the perception domain. \end{defn} 
\begin{rem} It should be emphasized that a temporal reality is defined only with respect to a certain temporal template. Some $\F$-observers may have several temporal templates, and some may have none. For example, since the $\C$-observer has no temporal templates, all realities of such an  observer are atemporal. It is easy to see that a temporal template and a global time uniquely determine the temporal order of $\mc{R}$. \end{rem} 
\begin{defn} For an $\F$-observer and a presence mode, $\overline{W}$, of his stable $\A$-phenomenon, the pullback, $\mc{T}_{\overline{W}} := \mc{T} \circ \sigma$, of the perceptible time $\mc{T}$ under the proper view $\sigma$ is referred to as the \emph{perceptible} \emph{time} \emph{of} $\overline{W}$. \end{defn} 
\begin{defn} For an $\F$-observer, let $\X$ be a stable  $\A$-phenomenon, $\mc{R} = (\A, \bo{R})$ a temporal reality, and $T$ a map $\mf{X} \to \F$. An ordered triple $\mb{X} = (\X, \mc{R}, T)$ is called a \emph{realization} (\emph{of} $\X$ \emph{in} $\mc{R}$) if the following diagram  commutes for each $\overline{W}$, 
\begin{displaymath} \label{BUNDLE} \xymatrix{\overline{W} \ar[d]_{j} \ar[r]^ \sigma \ar[dr]^ {\mc{T}_{\overline{W}}} & \mc{A}^{\bullet} \ar [d]^ {\mc{T}} \\ \mf{X} \ar[r]^ {T} & \F} \end{displaymath} where $j$ is the inclusion map. We refer to $\X$, $\mc{R}$ and $T$ as the \emph{underlying} \emph{phenomenon}, the \emph{background} \emph{reality} and the \emph{ambient} \emph{time} of the realization, respectively. \end{defn} 
\section{Observers} \label{Obs}
Due to the results of \cite{Tri95} and \cite{Tri07}, in the remainder of the paper we shall deal exclusively with the $\R$-observer, henceforth referred to simply as the \emph{observer}. If not mentioned explicitly, it is assumed in the following that the constructs under consideration always refer to the observer.
\begin{lem} Any rational paradigm of the observer is a unital algebra. \end{lem} 
\begin{proof} Since nonzero elements of $\A$ obey associative multiplication, nonassociativity can occur only in the permutations of $(\bo{ab})\bo{0}$, which is impossible since $\bo{b0} = \bo{0}, \forall \bo{b} \in \A$. Thus, $\A$ is associative, finite dimensional, and the identity of the motor domain is the identity of $\A$. Therefore it is unital. \end{proof} 
\begin{cor} The perception domain $\mc{A}^{\bullet}$ of a rational paradigm $\A$ is a lie group with respect to the multiplication of $\A$. \end{cor}
\begin{defn} For a rational paradigm $\A$, the linear manifold $\mc{A}$, canonically generated by the sensory domain $A$ is called the \emph{sensory} \emph{manifold} of $\A$, and each reflexor $a \in \mc{A}$ is called a \emph{viewpoint}. Viewpoints $a \in (\mc{A} \setminus)\mc{A}^{\bullet}$ are called \emph{(im)proper}. \end{defn} 
\begin{defn} For a rational paradigm $\A$, a reflexor $\bo{u} \in A$, and a proper viewpoint $a \in \mc{A}^{\bullet}$, the integral curve, through $a$, of the left invariant vector field generated by $\bo{u}$ is called a $(\bo{u}, a)$-\emph{vista}. \end{defn} 
\begin{rem} Intuitively, $(\bo{u}, a)$-vistas indicate naturally distinguished directions within the perception domain. \end{rem}
\begin{lem} The observer has a unique temporal template. \end{lem} \begin{proof} Indeed, there is a unique complete linear order on $\R$, namely the standard order $\leq$ (see, e.~g., \cite{BJ06}, p. 245). Thus, $(\R, \leq)$ is unique. \end{proof} 
\begin{rem} The previous result makes it unnecessary to mention the temporal order explicitly, and we use the simplified notation $\mc{R} = (\A, \mf{T})$ for a temporal reality of the observer. \end{rem} 
\begin{rem} It is shown in \cite{Tri95} that besides the home paradigm, the observer has exactly two (up to an $\R$-algebra isomorphism) consistent paradigms, namely the one dimensional $\R$-algebra of reals, $\R$, and the two dimensional $\R$-algebra of complex numbers, $\C$, both subalgebras of $\HH$. \end{rem}
\begin{defn} A reality is called \emph{robust} if it is temporal and there exists a principal metric, $\bo{\mc{T}}$, on the perception domain $\mc{A}^{\bullet}$ of the underlying paradigm with the perceptible time $\mc{T}$ as its generating function. Given a robust reality $\mc{R} = (\A, \mf{T})$, the structure field of $\mc{A}^{\bullet}$ is called the \emph{structure} \emph{(field)} of the reality, and we refer to $d\mc{T}$ and $\bo{\mc{T}}$ as the \emph{ether} \emph{(field)}, and the \emph{metric} of $\mc{R}$, respectively. The ordered pair $\mc{S} = (\mc{A}^{\bullet}, \bo{\mc{T}})$ is called the \emph{spacetime} of $\mc{R}$. \end{defn} 
\begin{defn} A  realization $\mb{X} = (\X, \mc{R}, T)$ is called \emph{robust} if the background reality is robust, and the perceptible part of the underlying phenomenon $\X$ is a principal $\mc{A}^{\bullet}$-bundle $(\mf{X})$, such that the propensity realm $\mf{X}$ is its total space and each proper view $\sigma : \overline{W} \to \mc{A}^{\bullet}$ is a fiber diffeomorphism. The dimensionality of $\mf{X}$ is referred to as the \emph{rank} of $\mb{X}$. For a presence mode, $\overline{W}$, of the underlying phenomenon the pullback $\bo{\mc{T}}_{\overline{W}}$, of the metric of the background reality under the proper view $\sigma$ is called the \emph{metric} of $\overline{W}$. The ordered pair $\mc{W} = (\overline{W}, \bo{\mc{T}}_{\overline{W}})$ is referred to as a \emph{(possible)} \emph{world} of $\mb{X}$. The bundle $(\mf{X})$ is called the \emph{monocosm} of the realization. \end{defn} 
\section{Dynamical systems} \label{DS}
\begin{defn} A  realization $\mb{X} = (\X, \mc{R}, T)$ is called a \emph{dynamical} \emph{system} if it is robust and its propensity realm is a riemannian manifold $\mc{X} = (\mf{X}, \bo{g})$. The riemannian metric $\bo{g}$ is referred to as the \emph{propensity} \emph{metric} of $\mb{X}$.  \end{defn} 
\begin{rem} There is a natural connection on the monocosm $(\mc{X})$ of a dynamical system: the horizontal space at any point $\psi$ is defined as the set of tangent vectors orthogonal to the world $\mc{W}_{\psi}$ with respect to the propensity metric. We refer to this connection as the \emph{fundamental} \emph{connection} of the dynamical system. \end{rem} 
\begin{defn} For a dynamical system $\mb{X} = (\X, \mc{R}, T)$ a \emph{perceptible} is a smooth map $f : \mc{X} \to \mc{A}^{\bullet}$. A \emph{temporal} \emph{evolution} of a dynamical system $\mb{X}$ is an integral curve of a vector field $\bo{f}_T$ on $\mc{X}$, called the \emph{temporal} \emph{evolution} \emph{vector} \emph{field} of $\mb{X}$, such that 
\begin{equation} \label{EVOL} (dT)(\bo{u}) = \bo{g}(\bo{f}_T, \bo{u}) , \quad \forall \bo{u} \in \mc{X}[\begin{smallmatrix} 1\\0 \end{smallmatrix}] , \end{equation} 
The propensity realm of a dynamical system is referred to as its \emph{state} \emph{space}, and propensity modes are called \emph{states}. A state $\psi$ such that the vector $\bo{f}_T(\psi)$ is vertical is called the \emph{proper} state of $\mb{X}$, and the possible world $\mc{W}_{\psi}$ is called an \emph{accessible} \emph{world}. \end{defn} 
\begin{rem} A perceptible is a smooth restriction of an attribute of the underlying phenomenon to the perception domain. Intuitively, a temporal evolution of a dynamical system is the motion of the observer's proper viewpoint across possible worlds of the system along its temporal evolution vector field. At each point of an evolution the observer encounters a possible world, a diffeomorphically perturbed copy of the spacetime of the background reality, which contains perceptible information about the system. \end{rem}
\begin{defn} For a dynamical system $\mb{X}$, an ordered triple $(f, \phi, \psi)$, where $f$ is a perceptible and $\phi$, $\psi$ are states, is called an $f$-\emph{observation}. The states $\phi$ and $\psi$ are called the \emph{initial} and the \emph{final} states, and the worlds $\mc{W}_{\phi}$ and $\mc{W}_{\psi}$ are called the \emph{source} and the \emph{target} worlds of the $f$-observation, respectively. The \emph{propensity} $\rho(\phi, \psi) \in \R \cup \{\infty \}$ of an $f$-observation $(f, \phi, \psi)$ is defined by 
\begin{equation} \rho(\phi, \psi) := \left \{ \begin{array} {r@{\quad : \quad}l} \infty & \mc{W}_{\phi} = \mc{W}_{\psi} \\ d^{-1} & \mc{W}_{\phi} \neq \mc{W}_{\psi} \end{array} \right. , \end{equation}
where $d$ is the length of the shortest geodesic between $\phi$ and $\psi$. \end{defn} 
\begin{rem} Intuitively, propensity roughly quantifies accessibility of the target world - the father away, the less accessible it is. \end{rem} 
\section{Cosmologies} \label{Cosm} 
\begin{rem} A left invariant vector field $\hat{\bo{u}}$ on $\mc{H}^{\circ}$, generated by a vector $\bo{u} \in H$ with the components $(u^{\beta})$ in a canonical basis $(\bo{i}_{\beta})$, assigns to each point $a \in \mc{H}^{\circ}$ with canonical coordinates $(w$, $x$, $y$, $z)$ a vector $\hat{\bo{u}}(a) \in T_{a}\mc{H}^{\circ}$ with the components $\hat{u}^{\beta}(a) = (\bo{a}\bo{u})^{\beta}$ in the basis $(\partial_w$, $\partial_x$, $\partial_y$, $\partial_z)(a)$ on $T_{a}\mc{H}^{\circ}$: 
\begin{multline} \label{LVFIELDS} 
\hat{u}^0(a) = wu^0 - xu^1 - yu^2 - zu^3 , \quad \hat{u}^1(a) = wu^1 + xu^0 + yu^3 - zu^2 , \\ 
\hat{u}^2(a) = wu^2 - xu^3 + yu^0 + zu^1 , \quad \hat{u}^3(a) = wu^3 + xu^2 - yu^1 + zu^0 . 
\end{multline} \end{rem}
\begin{defn}  A robust reality is called a \emph{cosmology} if its underlying paradigm is isomorphic to the home paradigm of the observer.  \end{defn} 
\begin{rem} As follows from Theorem $4.1$ of \cite{Tri07}, the choice of cosmologies of the observer is extremely limited. In fact, there is a unique, up to the functional variable $\mf{T}$, cosmology, $(\HH, \mf{T})$. Let us review its basic properties.  \end{rem}
\begin{enumerate} \item The perception domain $\mc{H}^{\circ}$ of the underlying paradigm $\HH$ is the lie group of nonzero quaternions with the $\R \times \mb{S}^3$ topology, the product of the real line and a three-sphere.  \item The spacetime $\mc{S} = (\mc{H}^{\circ}, \bo{\mc{T}})$ of the cosmology is a smooth four dimensional manifold with a closed FLRW metric. 
\item Given a canonical sensory basis $(\bo{i}_{\beta})$ and, for a differentiable function $R : \R \to \R \setminus \{0\}$, a system of natural spherical coordinates  $(\eta$, $\chi$, $\theta$, $\varphi)$ on $\mc{H}^{\circ}$, related to the canonical coordinates by 
\begin{multline*} 
w = R(\eta)\cos(\chi), \quad x = R(\eta)\sin(\chi)\sin(\theta)\cos(\varphi), \\
y = R(\eta)\sin(\chi)\sin(\theta)\sin(\varphi), \quad z = R(\eta)\sin(\chi)\cos(\theta) , 
\end{multline*} 
the metric has the following components in the spherical frame $(\partial^R_{\eta}$, $\partial^R_{\chi}$, $\partial^R_{\theta}$, $\partial^R_{\varphi})$  
\begin{multline} \label{MET} \mc{T}_{\alpha \beta} = diag(1, -\mf{a}^2, -\mf{a}^2{\sin^2\chi}, -\mf{a}^2{\sin^2\chi} {\sin^2\theta}) , \quad \mf{a}:= \sqrt{\mid \dot{\mc{T}} \mid} , \\ \text{with} \quad \quad R = \exp\int\frac{d\eta}{\genfrac{}{}{0pt}{3}{+}{-}\sqrt{\mid \dot{\mc{T}} \mid}} . \end{multline} 
\item The perceptible time $\mc{T}$ is a monotonous function of $\eta$. 
\item The structure field of the cosmology has the following components in the frame $(\partial^R_{\eta}$, $\partial^R_{\chi}$, $\partial^R_{\theta}$, $\partial^R_{\varphi})$:
\begin{multline*} \label{SF} \mc{H}^0_{\alpha \beta} = 
\begin{pmatrix} \lambda & 0 & 0 & 0 \\ 0 & -\lambda^{-1} & 0 & 0 \\ 0 & 0 & -\lambda^{-1}\sin^2\chi & 0 \\ 0 & 0 & 0 & -\lambda^{-1}\sin^2\chi\sin^2\theta \end{pmatrix},\\ 
\mc{H}^1_{\alpha \beta} = \begin{pmatrix} 0 & \lambda & 0 & 0 \\ \lambda & 0 & 0 & 0 \\ 0 & 0 & 0 & \sin^2\chi\sin\theta \\ 0 & 0 & 
-\sin^2\chi\sin\theta & 0 \end{pmatrix}, \\ \mc{H}^2_{\alpha \beta} = \begin{pmatrix} 
0 & 0 & \lambda & 0 \\ 0 & 0 & 0 & -\sin\theta \\ \lambda & 0 & 0 & 0 \\ 0 & \sin\theta & 0 & 0 \end{pmatrix},\ \mc{H}^3_{\alpha 
\beta} = \begin{pmatrix} 0 & 0 & 0 & \lambda \\ 0 & 0 & 1/\sin\theta & 0 \\ 0 & -1/\sin\theta & 0 & 0 \\ \lambda & 0 & 0 & 0 
\end{pmatrix} ,  \end{multline*} where $\lambda := \dot{R}/R$. 
\item The ether field of the cosmology has the following components in the frame $(\partial^R_{\eta}$, $\partial^R_{\chi}$, $\partial^R_{\theta}$, $\partial^R_{\varphi})$: 
\begin{displaymath} d\mc{T} = (\dot{\mc{T}}, 0, 0, 0). \end{displaymath} 
\item For a canonical sensory basis $(\bo{i}_{\beta})$ and the corresponding canonical coordinate system $(w, x, y, z)$, let $\bo{u}$ be a reflexor with the components $(u^\beta)$ in $(\bo{i}_{\beta})$, and $a$ - a proper viewpoint with coordinates $(\bar{w}, \bar{x}, \bar{y}, \bar{z})$. Then the $(\bo{u}, a)$-vista can be computed by solving the system of differential equations \eqref{LVFIELDS} with a parameter $t$: 
\begin{multline} w(t) = \exp(u^0t)(\frac{-u^1\bar{x} - u^2\bar{y}-u^3\bar{z}}{\omega}\sin{\omega t}+\bar{w}\cos{\omega t}) \\ 
x(t) = \exp(u^0t)(\frac{u^1\bar{w} - u^2\bar{z} + u^3\bar{y}}{\omega}\sin{\omega t}+\bar{x}\cos{\omega t}) \\ 
y(t) = \exp(u^0t)(\frac{u^1\bar{z} + u^2\bar{w} - u^3\bar{x}}{\omega}\sin{\omega t}+\bar{y}\cos{\omega t}) \\ 
z(t) = \exp(u^0t)(\frac{-u^1\bar{y} + u^2\bar{x} + u^3\bar{w}}{\omega}\sin{\omega t}+\bar{z}\cos{\omega t}) , \end{multline} 
where $\omega := \sqrt{(u^1)^2 + (u^2)^2 + (u^3)^2}$.
\end{enumerate} 
\begin{rem} The theory of the observer we have developed so far is left invariant (utilizing left invariant vector fields on perception domains). It is easy to show that the metric of the right invariant cosmology coincide with \eqref{MET}, but the $(\bo{u}, a)$-vistas are different. This can be used, in principle, by the observer to determine  the ``chirality'' of his contemporary reality. \end{rem} 
\section{Physical systems} \label{PHSYS} 
\begin{defn} A dynamical system $\mb{X} = (\X, \mc{R}, T)$ is called a \emph{physical} \emph{system} if $\mc{R}$ is a cosmology and the state space is a hyperk\"{a}hler manifold. \end{defn} 
\begin{defn} For a physical system $\mb{X} = (\X, \mc{R}, T)$, a quaternionic regular perceptible $f$ is called an \emph{observable}. For an observable $f$, we refer to its hyperhamiltonian vector field $\bo{f}$ as the $f$-\emph{field}. A state $\psi$ such that the vector $\bo{f}(\psi)$ is vertical is called an $f$-\emph{proper} state of $\mb{X}$. A possible world $\mc{W}$ is called $f$-\emph{(in)accessible} iff there is (not) an $f$-proper state $\psi$, such that $\mc{W} = \mc{W}_{\psi}$. For each $f$-proper state $\psi$ the value $f(\psi)$ is called a \emph{relative} \emph{perceptible} \emph{property} of $\mb{X}$. \end{defn} 
\begin{rem} The above definition generalizes the notions of GQM, where points at which the schr\"{o}dinger vector field of an observable becomes vertical, and the corresponding values of $f$ parametrize the eigenvectors and eigenvalues of $\widehat{\bo{F}}$, respectively  \cite{Sch96}. \end{rem}
\begin{defn} For a physical system $\mb{X} = (\X, \mc{R}, T)$, its \emph{hamiltonian} is an observable $h$ such that the temporal evolution vector field of $\mb{X}$ coincides with the $h$-field. \end{defn}
\begin{rem} A physical system can be thought of as a \emph{sufficiently} \emph{smooth} \emph{fine}-\emph{graining} of a cosmology. It is a natural generalization of the notion of a quantum system of GQM. \end{rem} 
\begin{defn} For a physical system, an $f$-observation $(f, \phi, \psi)$ is called \emph{successful} if $f$ is an observable, the propensity $\rho(\phi, \psi)$ exists, and the target world $\mc{W}_\psi$ is $f$-accessible, in which case we refer to $\mc{W}_{\psi}$ as the \emph{actual} \emph{world} of the $f$-observation. Otherwise the $f$-observation is called \emph{failed} (or \emph{unsuccessful}), and the world $\mc{W}_{\psi}$ is called \emph{virtual}. \end{defn} 
\begin{rem} A successful $f$-observation is also referred to as a \emph{measurement}. \end{rem}
\begin{defn} For a measurement $(f, \phi, \psi)$ its \emph{result} is an ordered pair $(\mc{W}_{\psi}, \mc{T}[{\psi}])$, where $\mc{W}_{\psi}$ is the actual world and $\mc{T}[{\psi}]$ is the level set of the perceptible time of $\mc{W}_{\psi}$ containing $\psi$. We refer to $\mc{T}[{\psi}]$ as the \emph{hypersurface} \emph{of} \emph{the} \emph{present}. The value $f(\psi)$ is called a \emph{relative} \emph{observable} \emph{property} of the physical system. \end{defn}
\begin{rem} Intuitively, the final state of an $f$-measurement marks the ``landing spot'' of the observer (or, more correctly, of his proper viewpoint) in a (target) world whose properties may differ, in a strictly defined sense, from the respective properties of the source world. Propensity carries information about the ``likelihood'' of the ``collapse'' of $\phi$ to $\psi$: the farther away $\psi$ is from $\phi$, the less likely it is that the final state is $\psi$. We are to think of a relative observable property as a generalization of a non-normalized eigenvalue of an observable in CQM. The main difference between the two notions is this: instead of a single real number it is an element of the quaternion algebra, which, given a canonical sensory basis, is an ordered quadruple of real numbers, the first of which can be used to represent the ``landing'' moment in the ambient time of the system, and the other three - to quantify various features of the system. Indeed, for a measurement to be a meaningful procedure, the time of its occurrence must be known to the observer. Even then, if a measurement of, say, a position of a particle in CQM results in a particular value of one of the coordinates, the position of the particle is still unknown unless the other two coordinates are known. \end{rem} 
\begin{defn} A physical system is called a \emph{hyperquantum} \emph{system} if its monocosm is a hyperquantum bundle $(\mc{V}_{\HH}^{\circ})$ over a quaternionic hilbert space $\V$, and the imaginary part $\mk{Im}(h)$ of its hamiltonian coincides with the restriction to $\mc{V}_{\HH}^{\circ}$ of the expectation of a quaternionic antihermitian operator $\widehat{\bo{H}}$ on $\V$. \end{defn} 
\begin{rem} It seems tempting to obtain quantum systems of GQM by demanding the existence of a canonical sensory basis in which the hamiltonian has a unique nonzero constituent $h_p$. However, this would not be quite correct because within our framework the underlying paradigm of such systems would correspond to the two dimensional consistent (boolean) paradigm $\C$, and their background reality is not a cosmology. In this sense quantum systems are a \emph{degenerate} case of hyperquantum systems: two out of four dimensions are collapsed in each possible world. \end{rem} 
\begin{defn} A dynamical system is called a \emph{quantum} \emph{system} if its background reality is $(\C, \mf{T})$, its monocosm is a quantum bundle $(\mc{V}^{\circ}_{\C})$ over a complex hilbert space $\V$, and the complex imaginary part of its hamiltonian coincides with the restriction to $\mc{V}^{\circ}_{\C}$ of the expectation of a complex antihermitian operator $\widehat{\bo{H}}$ on $\V$. \end{defn}
\begin{rem} It is a standard result in GQM that a schr\"{o}dinger evolution of a quantum system is also a hamiltonian evolution with the expectation of $\widehat{\bo{H}}$ as its generating function \cite{Sch96}, so the above definition is equivalent to the description of a quantum system in GQM. For an $f$-measurement $(f, \phi, \psi)$ of a quantum system the propensity $\rho(\phi, \psi)$ can be expressed in terms of probability of obtaining a particular result (\cite{BH01} and references therein). \end{rem} 
\begin{exmp} The \emph{Old} \emph{World} $\Upsilon$ . The monocosm of $\Upsilon$ is the \emph{degenerate} hyperquantum bundle with a single possible world, hence the source, target and actual worlds coincide for each measurement, and the ambient time of $\Upsilon$ is the perceptible time of the background cosmology. A temporal evolution of the system is orthogonal to the level sets of the perceptible time with respect to the propensity metric of $\Upsilon$. Since the evolution vector field is vertical everywhere, every state of the system is proper. For each measurement of $\Upsilon$ the metric of its actual world is the metric of the background cosmology.  In conventional terms this hyperquantum system seems to correspond to the classical coarse-grained view of the universe - an observer at rest relative to CMB in the spacetime of a spatially closed FLWR cosmology. \end{exmp}
\section{Summary} \label{SUM} As we mention in the introduction, the technical purpose of the paper is to provide \emph{formal} \emph{definitions} of observer related notions which are normally considered too ambiguous for constructive discussion within mainstream physics. Below is an informal overview of some of them. 
\par The \emph{observer} is represented by an \emph{existence} \emph{mode} of a \emph{boolean} \emph{experient}, and is capable of perceiving various \emph{realities}, each based on a \emph{paradigm}. In some realities the observer tends to distinguish \emph{dynamical} \emph{systems}, collections of experiences stable in a strictly defined sense. Dynamical systems spend most of their \emph{ambient} \emph{time} roaming their \emph{possible} \emph{worlds} according to the equation \eqref{EVOL} which is the main dynamical equation of the modification.  An \emph{observation} of a dynamical system perturbs its evolution resulting occasionally in a creation of an \emph{observable} \emph{property} of the system with respect to its \emph{actual} \emph{world}. For reasons that are beyond the scope of this theory, the contemporary \emph{operational} \emph{logic} of the observer seems to be bivalent boolean \emph{(Assertion 1)}, and hence the largest immediate environment conforming to this requirement corresponds to a \emph{robust} \emph{reality} of his \emph{home} \emph{paradigm} (a \emph{cosmology}). Then the kinematic axioms of GR follow: the spacetime of the cosmology is a smooth manifold, because it is a lie group, its dimensionality is indeed four, and it has a lorentzian metric of a very special type (closed FLRW). It is curious that the requirement of booleanity alone is sufficient, and bivalence follows (see \cite{Gol84}, p. 121). The internal mathematics of the $\HH$-universe, although boolean, is nonclassical: an important version of the axiom of choice fails in the topos $\mc{H}^{\circ}\bo{Set}$ (see \cite{Gol84}, p. 301), making the basic tool for obtaining, say, countable additivity of the Lebesgue measure unavailable to the observer. Since spacetime acquires a locally compact lie group structure, the observer can use the Haar measure whose properties are less dependent on the axiom of choice and can be described constructively (\cite{Car40}, \cite{Alf63}). The nonstandard integration over spacetime may have some application to the problem of apparently non-Newtonian behavior of large structures in contemporary observational astrophysics (see \cite{Mas07} and references therein).
\par \textbf{Acknowledgments}. The author is grateful to the reviewers whose comments helped reduce the volume and improve clarity of the exposition.

\end{document}